\documentclass[
    final,            % use final for the camera ready runs
  ]
  {aipproc}

\layoutstyle{6x9}

\usepackage{bm}

\newcommand{\nn}{\nonumber} 

\newcommand{\bea}{\begin{eqnarray}}
\newcommand{\eea}{\end{eqnarray}}

\newcommand{\mpi}{m_{\pi}}
\newcommand{\md}{m_{D}}

\newcommand{\mx}{m_{X}} 
\newcommand{\mdds}{\mu_{DD*}}
\newcommand{\vd}{v_D}

\newcommand{\mev}{\textrm{MeV}}

\newcommand{\Hb}{\bar{H}}

\newcommand{\bd}{ \bm{  D  } }
\newcommand{\bdbar}{ \bar{  \bm{ D } } }

\begin{document}

\title{$X(3872)$ in Effective Field Theory}


\author{S. Fleming}{
  address={Physics Department, University of Arizona, Tucson, AZ 85721}
}

\author{T. Mehen}{
  address={Physics Department, Duke University, Durham NC, 27708}
}

\begin{abstract}

If the $X(3872)$ resonance is a shallow boundstate of a the charm mesons
$D^{0} \bar D^{*0}$ and $D^{*0} \bar D^{0}$, it can be described by an effective theory of nonrelativistic $D$ mesons coupled to nonrelativistic pions (X-EFT).  In this talk, I give a brief overview of the $X(3872)$, followed by a short review of  X-EFT. I end my talk with results from calculations of the next-to-leading-order correction
to the partial decay width  $\Gamma[X\to D^0 \bar D^{0} \pi^0]$,
and the decay of $X(3872)$ to $P$-wave quarkonia.

\end{abstract}

\maketitle

%%%%%%%%%%%%%%%%%%%%%%%%%%%%%%%%%%%%%%%%%%%%
%% MAINMATTER
%%%%%%%%%%%%%%%%%%%%%%%%%%%%%%%%%%%%%%%%%%%%

\section{Introduction}

  The $X(3872)$~\cite{Choi:2003ue,Acosta:2003zx,Abazov:2004kp} is a novel charmonium state that lies very close to the  $D^0 \bar{D}^{*0}$ 
threshold. The closeness of the $X(3872)$ to this threshold $(-0.6 \pm 0.6\, {\rm MeV})$ has prompted numerous 
authors to suggest that it is a $C=1$ molecular bound state of $D^0  \bar{D}^{*0} + \bar{D}^0 D^{*0}$,
for reviews see Refs.~\cite{Voloshin:2006wf,Godfrey:2008nc}.
The strongest evidence for the molecular hypothesis is the large ratio of the
 branching fractions for the $X(3872)$ decays to $J/\psi$ plus two or three pions~\cite{Abe:2005ix} 
\bea\label{isoval}
\frac{{\rm Br}[X(3872)\to J/\psi \,\pi^+ \pi^- \pi^0]}
     {{\rm Br}[X(3872)\to J/\psi \, \pi^+ \pi^-]}
= 1.0 \pm 0.4 \pm 0.3 \, .
\eea 
The final states have opposite G-parity which implies that the $X(3872)$ does not have definite isospin. Measurements of the invariant mass distributions of the pions 
indicate that  these decays proceed through $J/\psi \,\rho$ for the 
$J/\psi \,\pi^+\, \pi^-$ final state, and through $J/\psi \, \omega$ for
the $J/\psi \,\pi^+ \,\pi^-\, \pi^0$ final state, suggesting that the $X(3872)$ couples to $I=0$ and $I=1$ channels with roughly equal strength. 
This is not possible for a conventional $c\bar{c}$ state. The observed branching ratio
can be understood if the $X(3872)$ is comprised primarily 
of neutral $D$ mesons, which is expected since the charged $D$-$D^*$ threshold
is far ($8.5 \,{\rm MeV}$) above the $X(3872)$ mass. There is, however, a caveat, as pointed out in Refs.~\cite{Suzuki:2005ha,DeFazio:2008xq} the ratio above could also be due to the kinematic suppression of the $J/\psi \, \omega$ mode, which leaves an opening for a charmonium interpretation of the $X(3872)$. Another possibility, suggested in 
Ref.~\cite{Gamermann:2007fi}, is that there are 
actually two nearly degenerate states
with opposite G-parity.

\section{X-EFT}

If the $X(3872)$ is a molecular state then the large distance behavior of the $D \bar{D}^*$ can be described by X-EFT~\cite{Fleming:2007rp,Fleming:2008yn}, an effective field theory which will be described below. 
The upper bound on the typical momentum
of the $D$ and $\bar D^*$ in the bound state is
$\gamma \equiv ( 2 \mdds E_X )^{1/2} \le 48 \, \mev$, 
where $\mdds$ is the reduced mass 
of the $D^0$ and $\bar D^{*0}$. For this binding momentum the 
typical velocity of the $D$ and $D^*$ is approximately 
$\vd \simeq ( E_X / 2 \mdds )^{1/2} \leq 0.02$, 
and both the $D$ and $D^*$ are non-relativistic. 

The pion degrees of freedom are also treated non-relativistically.  
The maximum energy of the pion emitted in the decay 
$X \to D^{0}  \bar D^{0} \pi^0$ is
\begin{eqnarray}
E_\pi = \frac{ 
\mx^2 - 4 \md^2 + \mpi^2}{ 2 \mx }= 142 \, \mev,   
\end{eqnarray} 
which is just 7 MeV above the $\pi^0$ mass at $134.98\, \mev$. 
The maximum pion momentum is approximately $44\, \mev$, which is comparable
to both the typical $D$-meson momentum, 
$p_D \sim \gamma \leq 48 \, \mev$,
and the momentum scale appearing
in the pion-exchange graph: $\mu \simeq 45\, \mev$.  Since the velocity of 
the pions is $v_\pi = p_\pi/m_\pi \leq 0.34$, a non-relativistic treatment 
of the pion fields is valid. 

The effective Lagrangian includes 
the charm mesons, the anti-charm mesons,  and the pion fields.  
We denote the fields that annihilate the 
$D^{*0}$, $\bar D^{*0}$, $D^{0}$, $\bar D^{0}$,
and $\pi^0$  as $\bd$, $\bdbar$, $D$, $\bar D$, and $\pi$, respectively. 
To the order we are working we will not need diagrams with charged pions and 
charged $D$ mesons so these are neglected in what follows. 
We construct an effective Lagrangian that is relevant for 
low-energy $S$-wave $DD^*$ scattering, 
where the initial and the final states are 
the $C=+$ superposition of 
$D^{0}\bar D^{*0}$ and $D^{*0}\bar D^{0}$:
\begin{eqnarray}
 |D D^*\rangle \equiv \frac{1}{\sqrt{2}}
 \left[
 |D^{0}\bar D^{*0} \rangle + | D^{*0}\bar D^{0} \rangle
 \right]. 
\label{dd*}
\end{eqnarray} 
An interpolating field with these quantum numbers will be used to calculate
the properties of the $X(3872)$.
We integrate out all momentum scales much larger than the momentum 
scale set by $p_D \sim p_\pi \sim \mu$. For $D$ mesons this corresponds 
to kinetic energy $\leq 1 \, \mev$,
for pions the kinetic energy is $\leq 7 \, \mev$. 
The hyperfine splitting $\Delta$  and $m_\pi$ 
should be treated as large compared to the typical energy scale in the
theory. 
The effective X-EFT Lagrangian is 
\begin{eqnarray}
 {\cal L}_{ } 
  &=& 
 \bd^{\dagger} \left(i\partial_0 + {\vec{\nabla}^2\over 2 m_{D^*}
		    }\right)\bd 
 +   D^\dagger \left(i\partial_0 + {\vec{\nabla}^2\over 2 m_{D} } \right) D 
 + \bdbar^{\dagger} \left(i\partial_0 + {\vec{\nabla}^2\over 2 m_{D^*}
		    }\right)\bdbar 
 \nonumber \\
 &+& 
 \bar D^\dagger \left(i\partial_0 + {\vec{\nabla}^2\over 2 m_{D}
		    }\right) \bar D 
 + \pi^\dagger \left(i\partial_0 + {\vec{\nabla}^2\over 2 m_{\pi}}
   + \delta\right) \pi
%%%
\nonumber \\
&+&  
\left(\frac{g}{\sqrt{2} f_\pi}\right) \,\frac{1}{\sqrt{ 2m_\pi } }
 \left( D \bd^\dagger \cdot \vec{\nabla}\pi  
   + \bar D^\dagger \bdbar \cdot \vec{\nabla}\pi^\dagger \right) + {\rm
 h.c.}
\nonumber \\
 &-& 
\frac{C_0}{2} \, \left(\bdbar D + \bd \bar D \right)^\dagger 
\cdot \left(\bdbar D + \bd \bar D \right) 
\nonumber \\
&+&   \frac{C_2 }{16} \, 
\left(\bdbar D + \bd \bar D \right)^\dagger 
\cdot \left(\bdbar ( \nabla)^2 D 
        + \bd (
         \nabla)^2 \bar D \right) + h.c.
\nonumber \\
%%%
&+&   \frac{B_1 }{\sqrt{2}}\frac{1}{\sqrt{2 m_\pi}} \left(\bdbar D + \bd \bar D \right)^\dagger \cdot D \bar{D} \vec{\nabla} \pi + h.c. + \cdots, 
\label{lag}
\end{eqnarray}
where 
$\delta = \Delta - m_\pi \simeq 7\, \mev$, $ \nabla$ is the operator $ \vec{\nabla}$ operating to the right minus $ \vec{\nabla}$ operating to the left, $\mu^2 = \Delta^2 - m_\pi^2 \approx 2 m_\pi \delta $, and $\cdots$ denotes higher-order interactions.
The pion decay constant is $f_\pi = 132 \,\mev$ with our choice of normalization.
Additional terms that couple the $X(3872)$ to charmonium states can be added to the lagrangian above. For example, the coupling of the $\chi_{cJ}$ to heavy mesons is given by
\bea\label{lagchi}
{\cal L}_\chi = i\frac{g_1}{2} {\rm Tr}[\chi^{\dagger \, i}  H_a \sigma^i \, \Hb_a] 
+ \frac{c_1}{2} {\rm Tr}[ \chi^{\dagger\, i} H_a \sigma^j \,\Hb_b] \epsilon_{ijk} A^k_{ab}+{\rm h.c.} \, .
\eea
Note the appearance of additional unknown couplings: $g_1$ and $c_1$.
\section{Results}

\begin{figure}[t]
\includegraphics[width=12cm]{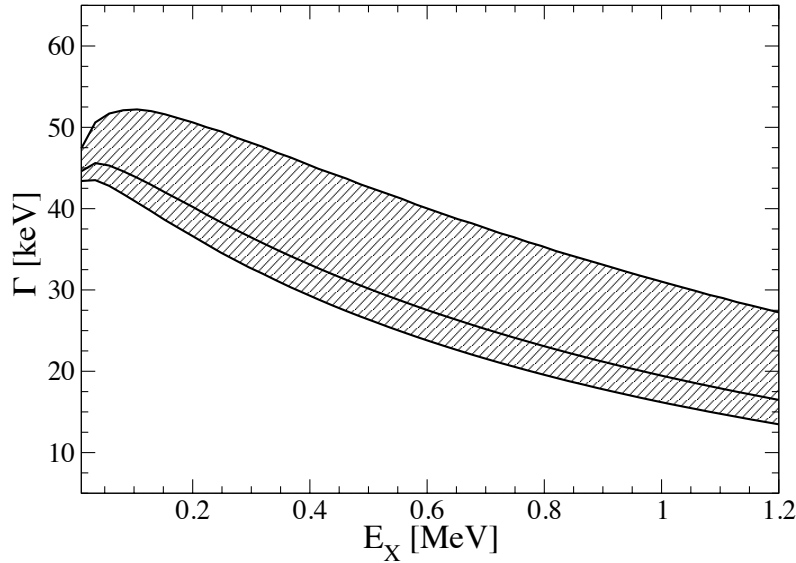}
\caption{ Decay rate for $X\to D^{0}\bar D^{0}\pi^0$ as a 
function of $E_X$. We use $g=0.6$. The central solid line 
corresponds to the 
LO prediction. The band is the 
result of the NLO calculation when the parameters $r_0$ and $\eta$
are varied in the ranges
$0 \leq r_0 \leq (100 \,{\rm MeV})^{-1}$ and $-1 \leq \eta \leq 1$.
\label{x-ddp}}
\end{figure}

Using the lagrangian given at the end of the last section it is straightforward to calculate the the next-to-leading-order correction
to the partial decay width  $\Gamma[X\to D^0 \bar D^{0} \pi^0]$, as well as the decays $X(3872) \to \chi_{cJ} \,(\pi^0,\pi\pi)$. In Fig.~\ref{x-ddp} we present the LO and NLO predictions for the decay rate for  $X\to D^{0}\bar D^{0}\pi^0$. Note, the LO decays $X(3872) \to D^0 \bar{D}^0 \pi^0$ was first calculated in Ref.~\cite{Voloshin:2003nt}, in addition the decay  $X(3872)\to D^0 \bar{D}^0 \gamma$ was calculated in Ref.\cite{Voloshin:2005rt}.
At NLO the decay rate depends on two new undetermined parameters, $r$ and $\eta$. In order to estimate their effect we vary these parameters within a reasonable range.

Using these results we can calculate the partial rates for $\Gamma[X(3872) \to \chi_{cJ}\,\pi^0]$. At LO
($c_1=0$) we find (denoting $\Gamma[X(3872) \to \chi_{cJ} \, \pi^0] \equiv \Gamma_J$)
\bea\label{lopred}
\Gamma_0:\Gamma_1:\Gamma_2::4.76:1.57:1 \, .
\eea
The energy dependence from the virtual pion propagators makes a significant change in the predictions
for the relative sizes of the partial rates. The NLO predictions depend on the ratio $c_1/g_1$ which is  undetermined.
The ratio $c_1/g_1$ has dimensions of inverse mass, when $c_1/g_1$ is varied from 
$(100 \,{\rm MeV})^{-1}$ to $(1000 \,{\rm MeV})^{-1}$, we find:
\bea\label{nlopred}
\Gamma_0:\Gamma_1:\Gamma_2&::&3.01:1.06:1 \qquad c_1/g_1=  (100 \,{\rm MeV})^{-1} \nn \\
\Gamma_0:\Gamma_1:\Gamma_2&::&3.49:1.20:1 \qquad c_1/g_1=  (300 \,{\rm MeV})^{-1} \nn \\
\Gamma_0:\Gamma_1:\Gamma_2&::&3.76:1.28:1 \qquad c_1/g_1=  (500 \,{\rm MeV})^{-1} \nn \\
\Gamma_0:\Gamma_1:\Gamma_2&::&4.11:1.38:1 \qquad c_1/g_1=  (1000 \,{\rm MeV})^{-1} \, .
\eea
For the largest values of $c_1/g_1$, we obtain predictions for the relative sizes of the partial rates
similar to those found in Ref.~\cite{Dubynskiy:2007tj}. As $c_1/g_1$ decreases, the predictions tend to those appearing
in Eq.~(\ref{lopred}). Experimental measurement of the relative sizes of the partial rates 
for decays to $\chi_{c,J}\,\pi^0$ can be used to determine $c_1/g_1$.

%%%%%%%%%%%%%%%%%%%%%%%%%%%%%%%%%%%%%%%%%%%%%%%%
%% BACKMATTER
%%%%%%%%%%%%%%%%%%%%%%%%%%%%%%%%%%%%%%%%%%%%%%%%

\begin{theacknowledgments}
 This work was supported in part by the Director, Office of Science, Office of Nuclear Physics, of the U.S. Department of Energy under grant number DE-FG02-06ER41449. 
\end{theacknowledgments}



\bibliographystyle{aipproc}   % if natbib is available


\end{document}